\documentclass[10pt,onecolumn,showpacs,amsmath,amssymb,superscriptaddress]{iopart}
\usepackage{iopams}  
\usepackage{bm}
\usepackage{ulem}
\usepackage[]{graphicx}
\usepackage[tight]{units}
\usepackage[]{nicefrac}
\usepackage{color}
\usepackage[colorlinks=true,citecolor=blue,linkcolor= blue,linkcolor=blue,urlcolor=blue]{hyperref}

\newcommand{\cip}{$\sigma_{_{\|}}$}
\newcommand{\cpp}{$\sigma_{_{\perp}}$}
\newcommand{\PFip}{$\text{PF}_{_{\|}}$}
\newcommand{\PFpp}{$\text{PF}_{_{\perp}}$}
\newcommand{\Sip}{$S_{_{\|}}$}
\newcommand{\Spp}{$S_{_{\perp}}$}
\newcommand{\mtpar}{m_{\|, \Delta_2}}
\newcommand{\mtper}{m_{\perp, \Delta_2}}
\newcommand{\mfpar}{m_{\|, \Delta_4}}
\newcommand{\mfper}{m_{\perp, \Delta_4}}

\begin{document}
\newcommand{\text}[1]{\mathrm{#1}}

\title[]{Effect of strain on the thermoelectric properties of silicon: An \textit{ab initio} study}
\author{N. F. Hinsche$^1$\footnote{Corresponding author: nicki.hinsche@physik.uni-halle.de}, I. Mertig$^1$,$^2$ and P. Zahn$^1$}
\address{$^1$ Institut f\"{u}r Physik, Martin-Luther-Universit\"{a}t Halle-Wittenberg, D-06099 Halle, Germany}
\address{$^2$ Max-Planck-Institut f\"{u}r Mikrostrukturphysik, Weinberg 2, D-06120 Halle, Germany}
\date{\today}

\begin{abstract}
 On the basis of detailed first-principles calculations the anisotropic thermoelectric transport 
 properties of biaxially strained silicon 
 were studied with focus on a possible enhancement of the powerfactor. Electron as well 
 as hole doping were examined in a broad doping and temperature range.
 In the low-temperature and low-doping regime an enhancement of the powerfactor 
 was obtained for compressive and tensile strain in the electron-doped case and for compressive strain 
 in the hole-doped case. In the thermoelectrically more important high-temperature and high-
 doping regime a slight enhancement of the powerfactor was only found for the hole-doped 
 case under small biaxial tensile strain. The results are discussed in terms of band-structure effects. 
 An analytical model is presented to understand the 
fact that the thermopower decreases if degenerate bands are energetically lifted due to a strain-
induced redistribution of states.
\end{abstract}

\submitto{\JPCM}

\maketitle


\section{Introduction}
Thermoelectrics convert heat into electric current, and vice versa. Known for more than 60
years, thermoelectrics attract currently attention 
\cite{Snyder:2008p240,Sales:2002p6580,Majumdar:2004p6568,Bottner:2006p2812}. 
With nearly 90 per cent of the 
worlds power being generated by low efficient heat engines, thermoelectric modules could 
potentially convert parts of this wasted heat into electricity. Their conversion efficiency can be stated
by the figure of merit
\begin{equation}
ZT=\frac{\sigma S^{2}}{\kappa_{el} + \kappa_{ph}} T,
\label{eq1}
\end{equation}
where $\sigma$ is the electrical conductivity, $S$ the thermopower, $\kappa_{el}$  and 
$\kappa_{ph}$ are the electronic and phononic contribution to the thermal conductivity, respectively. 
The numerator of Eq.~\ref{eq1} 
is called powerfactor $\text{PF}=\sigma S^{2}$.

While $ZT>1$ was challenging to be reached in the last decades, nowadays nanostructured 
thermoelectrics enable even larger values of $ZT$ \cite{Venkatasubramanian:2001p114,Harman:2002p5345,Dresselhaus:2007p2775}. 
Unfortunately, those materials are often based on 
environmentally unfriendly lead, tellurium or selenium compounds and are therefore hard to integrate in 
semiconductor electronics.

Silicon, the cradle of modern semiconductor electronics, is nonpolluting, readily available, cheap and 
perfectly integrated in present electronics infrastructure. While silicon has been 
stated as inefficient thermoelectric in the past due to his enormous thermal conductivity \cite{Vining:2008p9416}, 
recent experiments and theory revealed that nanostructuring could lead to thermoelectric efficiencies comparable 
to state of the art commercial thermoelectric materials \cite{Hochbaum:2008p6569,Boukai:2008p14967,Lee:1997p1545,Bux:2009p14985,Hao:2011p14983}.

As in nanostructured materials mechanical strain plays an 
important role, this paper will focus on the influence of biaxial strain on the electronic thermoelectric 
transport properties of bulk silicon , which might occur in rolled up and layered Si heterostructures \cite{Baykan:2011p14974,
Prinz:2006p8835,Schmidt:2001p7557,Cho:2006p8830}. It is well known, that similar strain physics, 
e.g. band splitting and band deformation are expected for strained three dimensional systems, as well as for 
one and two dimensional silicon devices in complex strain states. A comprehensive overview on this can be found in ref.~\cite{Baykan:2011p14974}. In detail 
these common strain physics affect the electronic carrier transport depending
on the dimensionality, temperature and doping density.
The question to be answered in this paper 
will be whether tensile or compressive strain will lead to an enhancement or reduction of 
the powerfactor in silicon under a certain doping and temperature environment.
Beside our interest in the high temperature 
thermoelectric application of strained silicon we want to emphasize the possible importance of our results in the 
low temperature regime 
for the metal-oxide-semiconductor device community, where the knowledge of the thermoelectric properties 
of silicon under strain could help to understand parasitic effects in these devices. In the low doping regime at low temperature, 
an enhancement of one part of 
the powerfactor, namely the electrical conductivity, under externally applied strain was found and heavily investigated in the last decades.\cite{Ieong:2004p15185,Fischetti:2002p15225,Baykan:2011p14974}
For this purpose the paper will be organized as follows. In section \ref{method} we introduce our first principle 
electronic structure calculations based on density functional theory and the transport calculations 
based on the solution of the linearized Boltzmann equation. With this knowledge we start the discussion of the thermoelectric 
transport properties of unstrained bulk silicon (Sec.~\ref{transport-unstrained}) and present afterwards the influence 
of biaxial strain on the electron or hole doped case of silicon in section \ref{transport-strain}. In the last paragraphs
\ref{optpf} and \ref{ZT} the optimal powerfactor under strain due to variation of doping is determined and analyzed, while 
estimations of the possible figure of merit are given.

\section{\label{method} Methodology}
Our approach is based on two ingredients: first principles density functional theory calculations (DFT), 
as implemented in the \textsc{QuantumEspresso} package \cite{Giannozzi:2009p14969} and an in-house developed Boltzmann 
transport code \cite{Mertig:1999p12776} to calculate the thermoelectric transport properties.

\subsection{Electronic structure}

In a first step the band structure of the strained and unstrained Si was calculated using 
the general gradient approximation (GGA) and the PBE exchange correlation functional  
\cite{Perdew:1981p14970}. Fully relativistic and norm-conserving pseudo potentials \cite{Corso:2005p8612} were 
used to treat the spin-orbit splitting of the Si valence bands in a correct way. The calculations were performed with the experimental 
lattice constant $ a=5.434~\AA $ for a face-centered tetragonal eight atom unit cell. 
The in-plane biaxial strain is simulated by changing the \nicefrac{c}{a} ratio but keeping the cell volume constant.
Throughout the 
paper the biaxial strain will be given in units of the relative change 
of the in-plane lattice constant as $\nicefrac{\Delta a}{a_{0}}=\nicefrac{a}{a_{0}}-1$. 
That means tensile strain considers changes $\nicefrac{\Delta a}{a_{0}}>0$, while compressive strain 
means $\nicefrac{\Delta a}{a_{0}}<0$.

As expected, our DFT calculations underestimate the size of the band gap at zero temperature and do not reproduce the 
temperature dependence of the gap. 
For this purpose we included a temperature-dependent scissor operator~\cite{Godby:1988p14795}, so that
the strain- and temperature-dependent energy gap $E_g$ in \unit[]{eV} becomes 
\begin{eqnarray}
\label{teg}
E_{g}(T,\frac{\Delta a}{a_0}) &=& E_{g}(T=0,\frac{\Delta a}{a_0}) + U_{\text{GGA}} \\ \nonumber
&-& \frac{4.73 \cdot10^{-4}\cdot T^{2}}{T+636},
\end{eqnarray}
where $E_{g}(T=0,\frac{\Delta a}{a_0})$ is the zero temperature gap obtained by our self-consistent DFT calculations, $U_{\text{GGA}}=\unit[0.57]{eV}$ 
is a static correction to fulfill the experimental low temperature gap. The third part of equation \ref{teg} is the correction of the 
temperature dependence of the band gap \cite{boernstein} in a wide temperature range with T given in K.
\subsection{Boltzmann transport}
With the converged results from the first step we are now able to obtain the thermoelectric transport properties 
by solving the linearized Boltzmann equation in relaxation time approximation (RTA) \cite{Mertig:1999p12776}. 
Boltzmann transport calculations for thermoelectrics have been carried out for quite a long time and show 
reliable results for metals~\cite{Vojta:1992p1395,Thonhauser:2004p14960,Yang:2008p11828,Barth:2010p15131} as well as for wide- and narrow gap semiconductors~\cite{Singh:2010p14285,Parker:2010p13171,May:2009p14962,Lee:2011p14982}.
Within in here the relaxation time is assumed to be constant with respect to wave vector k and energy on the scale of $k_{B}T$. 
This assumption is widely used for metals and doped semiconductors. The constant relaxation time 
is a big advantage for the calculation of the thermopower $S$ without any adjustable parameter, 
while lacking any doping or temperature dependence of $\tau$. For unstrained silicon, doping dependent relaxation times 
in the order of \unit[15-150]{fs} for electron doping and \unit[6-65]{fs} for hole doping could be estimated 
from experiments in ref.~\cite{canali77}. To concentrate on the bandstructure effects we assume the 
relaxation time does not depend on strain, while 
it was shown, that under strain the dominant scattering
process varies: For unstrained Si, the room-temperature scattering
is dominated by optical phonons, i.e., intervalley scattering,
whereas for strained Si, this scattering process is less efficient \cite{Dziekan:2007p1770,Roldan}.

The temperature- and doping-dependent thermopower in- and cross-plane is defined as
\begin{eqnarray}
S_{_{\perp, \|}}=\frac{1} {eT} \frac{\mathcal{L}_{\perp, \|}^{(1)}(\mu,T)} {\mathcal{L}_{\perp, \|}^{(0)}(\mu,T)}
\label{Seeb},
\end{eqnarray}
where
\begin{eqnarray}
& \mathcal{L}_{\perp, \|}^{(n)}(\mu, T)= \nonumber \\
&\frac{\tau}{(2\pi)^3} \sum \limits_{\nu} \int\ d^3k \left( v^{\nu}_{k,(\perp, \|)}\right)^2 (E^{\nu}_k-\mu)^{n}\left( -\frac{\partial f_{(\mu,T)}}{\partial E} \right)_{E=E^{\nu}_k} \nonumber
\\
\label{Tcoeff}
\end{eqnarray}
is the transport distribution function as termed by Mahan and Sofo \cite{Mahan:1996p508}, for given 
chemical potential $\mu$ at temperature $T$ and carrier concentration $N$ determined by an integration 
over the density of states $n(E)$
\begin{eqnarray}
N=\int \limits_{\mu-\Delta E}^{\text{VB}^{max}} \text{d}E \,  n(E) [f_{(\mu,T)}-1]+
\int \limits_{\text{CB}^{min}}^{\mu+\Delta E} \text{d}E \, n(E) f_{(\mu,T)}
\label{Dop},
\end{eqnarray}

where $\text{CB}^{min}$ is the conduction band minimum and $\text{VB}^{max}$ is the 
valence band maximum. 
As can be seen straight forward from eqs.~\ref{Seeb} and \ref{Tcoeff} 
the electrical conductivity $\sigma$ is then given by
\begin{eqnarray}
\sigma_{_{\perp, \|}}=2e^2 \mathcal{L}_{\perp, \|}^{(0)}(\mu, T)
\label{Sigma}.
\end{eqnarray}

The Fermi-surface integration, which is requested in equation \ref{Tcoeff}, is performed within 
an extended tetrahedron method \cite{Lehmann:1972p14972,Zahn:1995p14971,Mertig:1987p5922} interpolating the 
calculated Eigenvalues $E_k^{\nu}$ on a mesh of at least 13950 k points in the irreducible 
part of the Brillouin zone. $\mathcal{L}_{\perp, \|}^{(0)}(E, T=0)$ was determined 
on a fine energy mesh with a step width of $\unit[1]{meV}$. Convergence tests confirmed that the calculation 
of the electrical conductivity and the thermopower at a given temperature via equations \ref{Sigma} and \ref{Seeb} 
requires $\mathcal{L}_{\perp, \|}^{(0)}(E, T=0)$ for a quite large range. 
Convergence of the integrals \ref{Tcoeff} and \ref{Dop} was achieved with an adaptive integration method 
for $2\Delta E \geq 20k_b T$ in the limit of low carrier concentrations $N \leq \unit[1\cdot 10^{14}]{cm^{-3}}$.

\section{Thermoelectric transport}
\subsection{\label{transport-unstrained} Unstrained case}

\begin{figure}[t]
\centering
\includegraphics[width=0.88\textwidth]{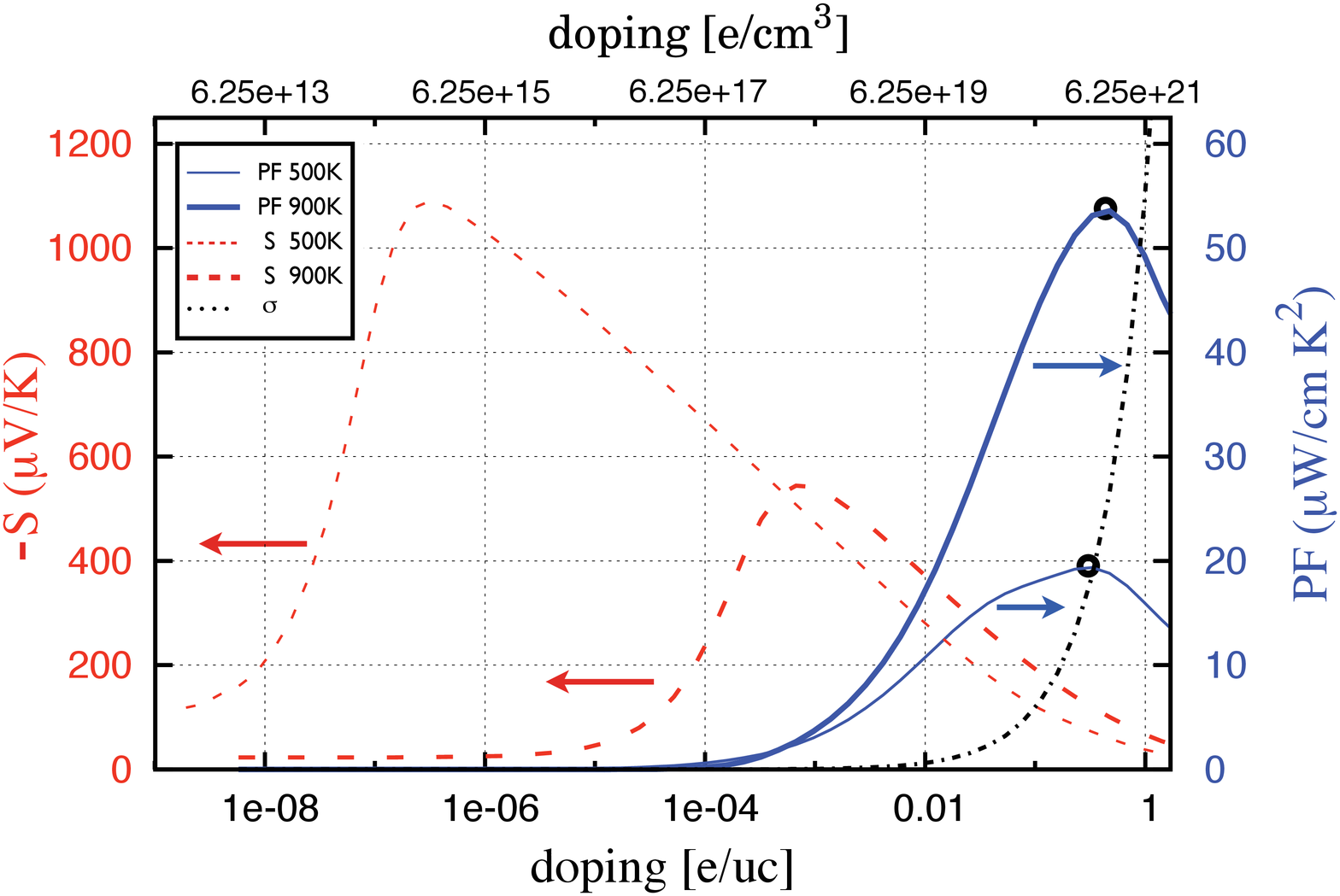}
\caption{\label{fig:1}(color online) In-plane thermopower (red dashed lines, refer to left scale, red units) and powerfactor  
(blue solid lines, refer to right scale, blue units) for unstrained, electron doped silicon in dependence 
on the doping level. Furthermore the doping dependence of the electrical conductivity is given as a 
dashed dotted line in arb. units. The maxima of the powerfactor are marked by black open circles.}
\label{fig:unstrained}  
\end{figure}

Since the thermoelectric transport properties of the strained silicon will always be 
discussed in comparison with the unstrained case, we will provide an insight into the transport 
properties of the unstrained silicon, first. In figure \ref{fig:unstrained}, the thermopower and the powerfactor 
are shown for two different temperatures (500K and 900K) in a wide doping range. The 
qualitative behavior of the electrical conductivity is indicated by the dashed-dotted line to 
emphasize the trend of the resulting powerfactor. The picture is well known for the 
interrelation of electronic transport and the 
thermoelectric properties of semiconductors \cite{Ioffe:1960,Snyder:2008p240}. 
At constant temperature the thermopower (cf. red dashed lines in figure \ref{fig:unstrained}) decreases at very low and 
higher doping level and reaches a maximum inbetween \cite{Ioffe:1960}.

As can be seen from figure \ref{fig:unstrained} the thermopower reaches 
a maximum around an electron carrier concentration of $\unit[1\cdot 10^{15}]{cm^{-3}}$ at 500K, while 
the maximum at 900K is shifted to a larger doping level of $\unit[1\cdot 10^{18}]{cm^{-3}}$. 
Beside that, the maximum of the more relevant powerfactor (cf. blue solid lines in figure \ref{fig:unstrained}, 
optimal values indicated by black open circles) 
is shifted to huge electron carrier concentrations of about $\unit[1\cdot 10^{21}]{cm^{-3}}$. 
This is determined by the linear increase of the electrical conductivity with increasing 
charge carrier concentration. 
Obviously, there is not much space to optimize the powerfactor with respect to temperature 
and charge carrier concentration for unstrained silicon. 
We will focus on this optimization in more detail in subsection \ref{optpf}.
\begin{figure}[t]
\centering
\includegraphics[width=0.50\textwidth]{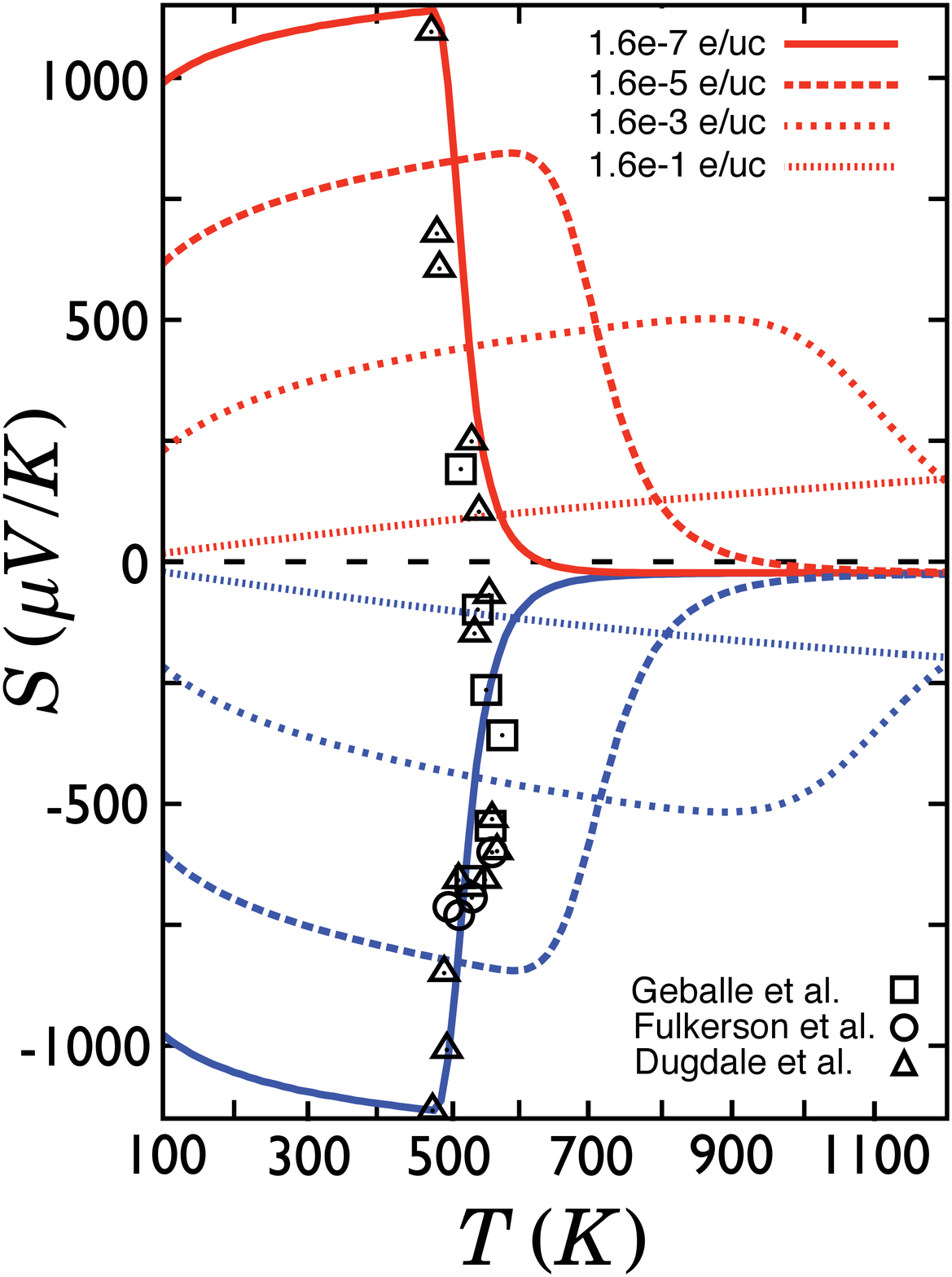}
\caption{\label{fig:2}(color online) Thermopower for unstrained silicon in 
dependence on doping and temperature. Electron doping refers to the blue lines
 in the lower part of the figure, while red lines refer to hole 
doping and positive values of the thermopower.
Experimental data (squares, circles and triangles) from ref.~\cite{Geballe:1955p9278, Fulkerson:1968p9375, Dugdale:1977} 
are given for comparison.}
\label{fig:seebeck}  
\end{figure}
The complex dependence of the thermopower on temperature and doping is shown in figure \ref{fig:seebeck} for 
electron and hole doped unstrained silicon at different doping concentrations.
For low temperatures and doping levels the thermopower reaches values of $\unit[1000]{\mu V/K}$ 
and above, which is caused by the location of the chemical potential near the band edges. The denominator of equation \ref{Seeb}, 
proportional to the electrical conductivity, is small, while the nominator is large, the thermopower gets maximal. At fixed 
charge carrier concentration the position of the chemical potential is shifted towards the middle 
of the gap with increasing temperature. The nominator in equation \ref{Seeb} decreases, because it is mainly determined by the opposite contributions of 
the tails of the derivative of the Fermi-Dirac distribution function with respect to the valence and conduction band (equation \ref{Tcoeff}, n=1). 
At a distinct temperature of about $\unit[500]{K}$ the thermopower rapidly vanishes. At this temperature the 
electronic transport enters into the bipolar intrinsic regime. To emphasize the relevance of our calculations experimental results 
for merely pure silicon in the intrinsic transport regime are added in figure \ref{fig:seebeck}. We want to point out, that the 
thermopower does not vanish at all, but converges to small negative values for electron, as well as for hole doping. 
At large charge carrier concentrations of about $\unit[0.16]{e/uc}$ (dotted lines in figure \ref{fig:seebeck}), where 
the powerfactor becomes large, the thermopower grows linearly with temperature upto values of $\unit[150]{\mu V/K}$ at $\unit[1000]{K}$. 
In the heavy doped regime the chemical potential is located deeply in the bands and 
equation \ref{Seeb} qualitatively coincides with the well known \textsc{Mott} relation $S \propto \frac{\text{d} \ln \sigma(E)}{\text{d}E}|_{E=\mu} $
for the thermopower in RTA \cite{Cutler:1969p3655}.

\subsection{\label{transport-strain} Optimization of the powerfactor by strain}
\subsubsection{\label{} Electron doping}

Having provided a general view on the thermoelectric transport properties of unstrained silicon 
above, we will now focus on the effect of biaxial strain on those properties. 
The results are presented in comparison to the unstrained case, starting with the 
electron doped case, followed by the hole doped case in subsection \ref{strain-hole}.

\begin{figure}[t]
\centering
\includegraphics[width=0.68\textwidth]{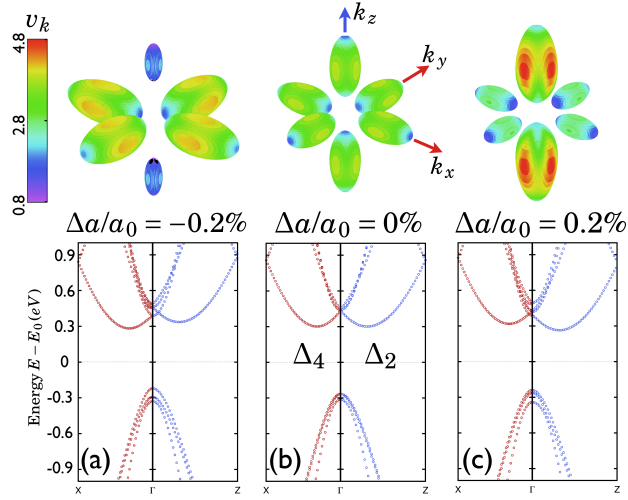}
\caption{\label{fig:3}(color online) Fermi surfaces of electron doped silicon under 
compressive strain (left), no strain (middle) and tensile strain (right). On the Fermi surfaces 
the absolute value of the group velocities are plotted in units of $\unit[0.08\cdot 10^{6}]{m/s}$. 
As reference the band structure on two high symmetry lines is given below. 
The doping corresponds to additionally $0.01$ electrons per unit cell which causes carrier 
densities of $\unit[6.25\cdot 10^{19}]{cm^{-3}}$.}
\label{fig:fermisur}  
\end{figure}

\begin{figure}[t]
\centering
\includegraphics[width=0.78\textwidth]{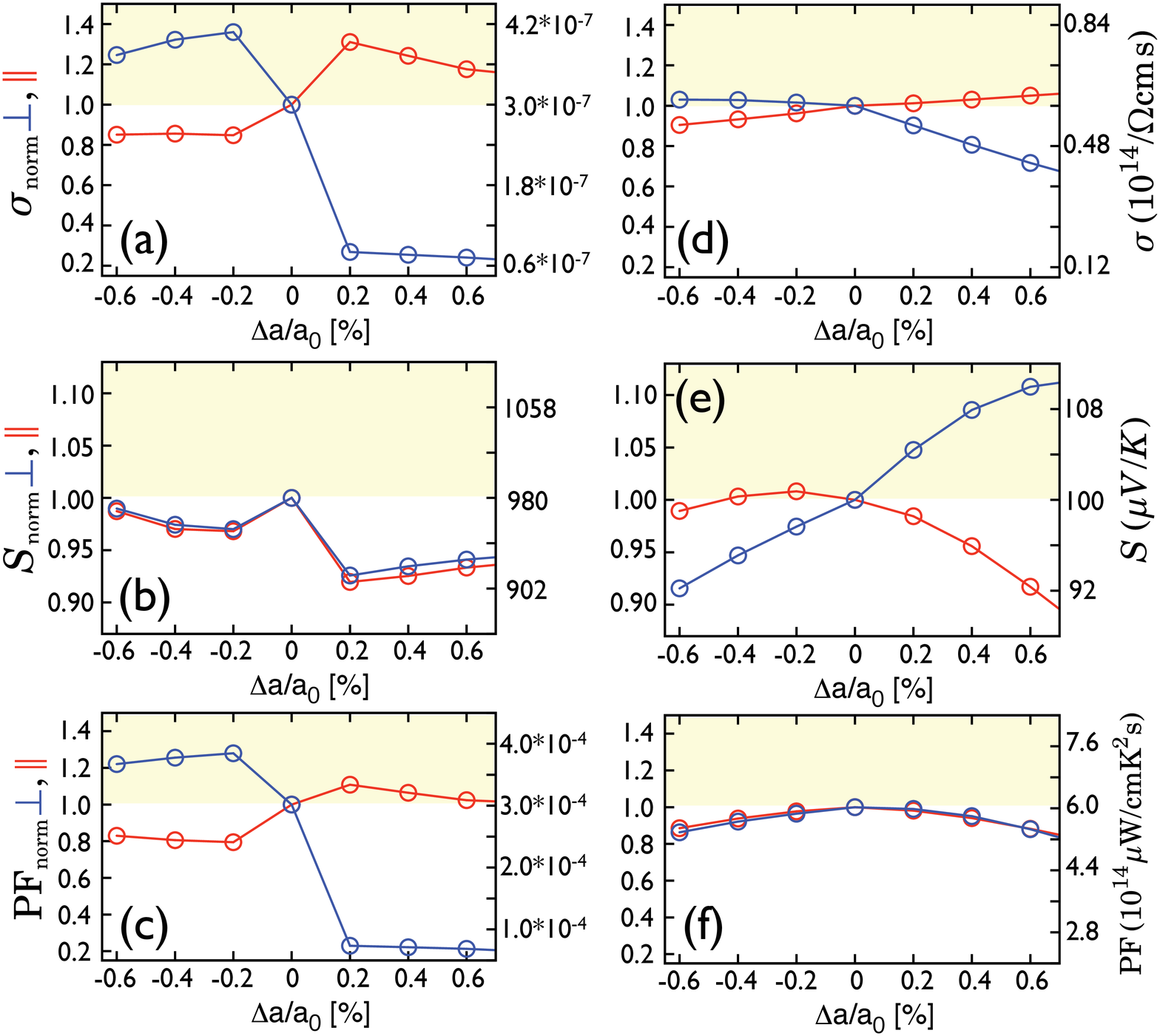}
\caption{\label{fig:4}(color online) Anisotropic thermoelectric transport properties for fixed temperature 
and electron doping concentrations in dependence on compressive and tensile strain. Left panels (a)-(c)) 
correspond to an electron doping of $\unit[1.6\cdot 10^{-7}]{e/uc}$ 
at a temperature of 100~K, while right panels refer to an electron doping of \unit[0.48]{e/uc} 
at a temperature of 900~K. On the left axis of each figure the relative value compared to the 
unstrained case is shown, while on the right axis the absolute values are given. Electrical conductivity ((a),(d))
 and powerfactor ((c),(f)) are presented in units of the relaxation time $\tau$.}
\label{fig:ndop}  
\end{figure}

In strain-free bulk silicon, as introduced in section \ref{transport-unstrained}, the 
conduction-band minimum (CBM) consists of six equivalent valleys on the $\Gamma$-X high symmetry line 
as shown in the middle panel of figure \ref{fig:3}(b). The Fermi surface
pockets corresponding to these valleys are shown on top, with the absolute value of the 
carriers group velocity, $v_k$ entering equation \ref{Tcoeff}, plotted in color on the surface. The color code gives $v_k$ 
in units of $\unit[0.08\cdot 10^{6}]{m/s}$. By applying biaxial in-plane strain, 
the six CBM valleys are energetically split into two groups. Four degenerate in-plane $\Delta_{4}$ valleys and 
two degenerate cross-plane $\Delta_{2}$ valleys (see figure \ref{fig:fermisur} (a),(c)). While the number of electrons 
is fixed, the different sizes of the ellipsoidal pockets is caused by a change of occupation numbers under strain. 
The color code indicates the overall smaller Fermi velocities on the small pockets and in particular on the principal axis of the 
pockets, whereas larger velocities are found for states propagating perpendicular to the principal pocket axis. 
In figure \ref{fig:ndop} the thermoelectric transport properties 
of biaxial strained silicon for two fixed electron doping regimes 
are shown. First, we consider the left column (figures \ref{fig:ndop}(a)-(c)), which shows the electrical 
conductivity $\sigma$, the thermopower $S$ and the resulting powerfactor $\text{PF}=\sigma S^2$ 
for a low electron doping of $\unit[1.6\cdot 10^{-7}]{e/uc}$ and a low temperature of \unit[100]{K}.
For this doping and temperature value a large enhancement of up to 35\% of the electrical conductivity is obtained 
for the in-plane component \cip at low tensile strain and for the cross-plane component \cpp at small
compressive strain. \cpp drops noticeable under small tensile 
strain up to 30\% of the unstrained case, while \cip experiences a slight drop down to 83\%  under small 
compressive strain, respectively.

With the conduction bands of silicon behaving parabolically near the band edges, the calculated transport 
properties can be understood for small charge carrier concentrations and low temperatures in terms of 
effective masses and occupation number redistributions (see \cite{Dziekan:2007p1770}. 
With increasing tensile strain the $\Delta_{4}$ bands lift up and the occupied states
from the higher bands are transferred to the lowered $\Delta_{2}$ bands 
(comp. figure \ref{fig:fermisur}(c)). At a certain tensile strain the $\Delta_{4}$ pockets are completely 
depleted and solely the maximally occupied $\Delta_{2}$ states contribute to the 
transport. 
In simplified consideration one can estimate the relative change in the electronic conductivity 
from the relative change in the effective electron mass. As reported earlier \cite{Dziekan:2007p1770} it is 
$\mtper=\mfper=0.205$ and $\mtpar=\mfpar=0.926$, 
whereas the masses are in units of the free electron rest mass. For the in-plane component \cip of the electrical 
conductivity at sufficient tensile strain only the lowered $\Delta_{2}$ bands contribute with their perpendicular mass $\mtper$. 
With the notation
\begin{equation}
\frac{1}{m_0} = \frac{1}{6} \left( \frac{2}{\mfpar} + \frac{2}{\mfper}  + \frac{2}{\mtper} \right) = \frac{1}{0.277},
\end{equation} 
the normalized asymptotic value becomes: 
\begin{equation}
\frac{1}{2} \left( \frac{2}{\mtper} \right) \cdot m_0 = 1.35
\end{equation}
For the in-plane component \cip of the electrical 
conductivity at compressive strain only the four pockets of the lowered $\Delta_{4}$ bands contribute 
equally with their parallel and perpendicular mass:
\begin{equation}
\frac{1}{4} \left( \frac{2}{\mfpar} + \frac{2}{\mfper} \right) \cdot m_0 = 0.83
.\end{equation}
For the cross-plane conductivity \cpp at tensile strain it is
\begin{equation}
\frac{1}{2} \left( \frac{2}{\mtpar} \right) \cdot m_0 = 0.30
,\end{equation} 
whereas at compressive strain \cpp becomes
\begin{equation}
\frac{1}{2} \left( \frac{2}{\mtper} \right) \cdot m_0 = 1.35
.\end{equation}
Since the powerfactor is composed out of $\sigma$ and $S$ we analyze the 
influence of strain on the thermopower as well. In the low temperature and doping regime 
(figure \ref{fig:ndop} left panel) no enhancement of thermopower at either compressive or tensile strain could be 
found. It can bee seen, that for tensile strain the thermopower decreases by about 10\%, while for compressive 
strain a drop of about 5\% is found. The difference between the in-plane and cross-plane thermopower is 
marginal. In terms of \textsc{Mott}s formula \cite{Cutler:1969p3655} it means that the energy dependence of 
$\mathcal{L}_{\perp, \|}^{(0)}(E, T)$ and $\mathcal{L}_{\perp, \|}^{(1)}(E, T)$ is almost the same. Nevertheless it is interesting to understand why 
the thermopower of silicon is reduced under strain and why the decay changes for tensile and compressive strain.

\begin{figure}
\centering
\includegraphics[width=0.68\textwidth]{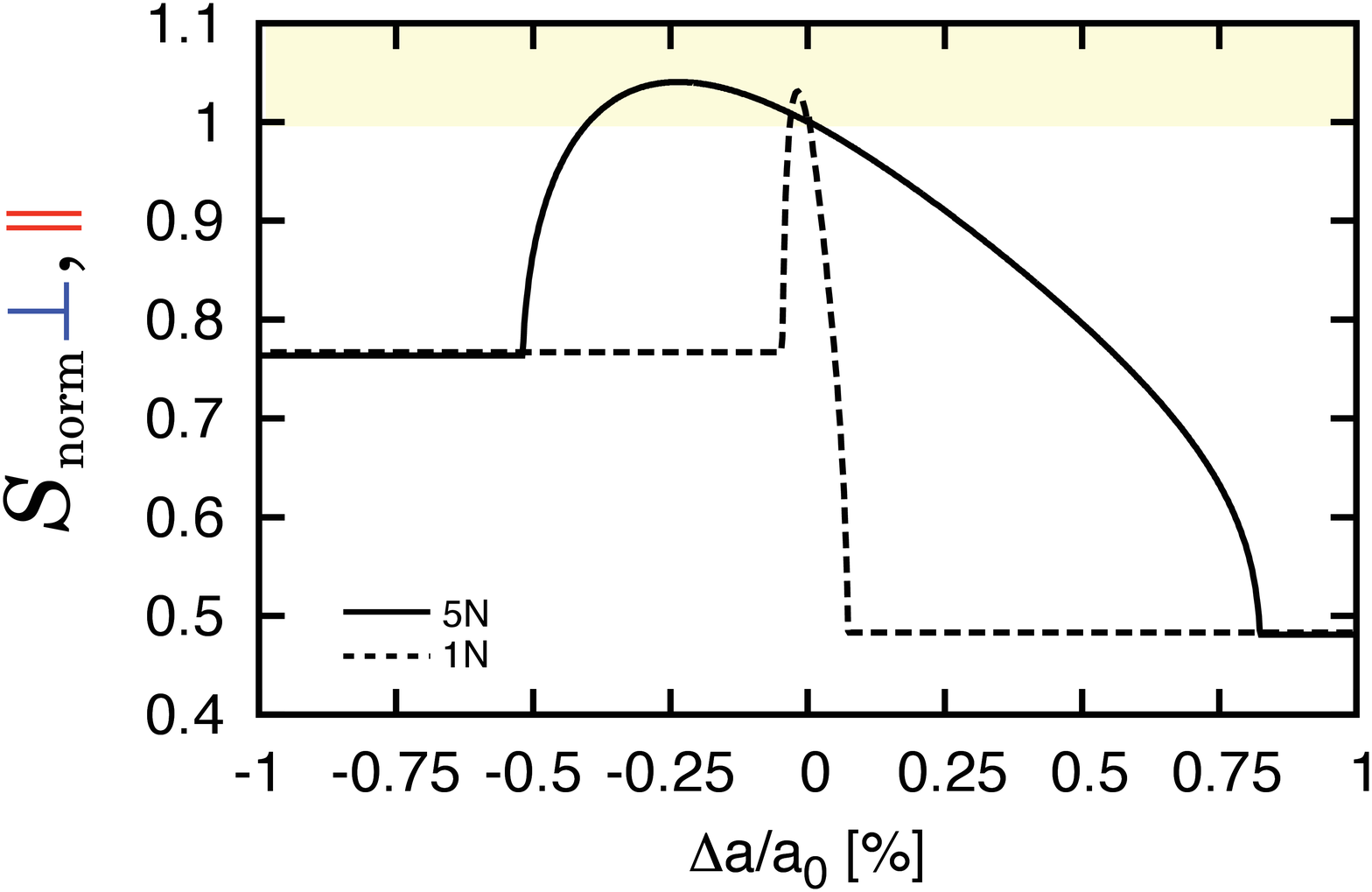}
\caption{Analytical dependence of the thermopower on biaxial strain for different electron charge carrier concentrations, 
small charge carrier concentration (dashes lines) and increased charge carrier concentration by a 
factor five (solid lines).}
\label{fig:anaS}
\end{figure}

For qualitative understanding of our \textit{ab initio} results we apply a free electron model to discuss the thermopowers behavior 
on biaxial strain. The strain dependent electrical conductivity at zero temperature was modeled as proposed 
by \cite{Dziekan:2007p1770} and then 
the thermopower was calculated by \textsc{Mott}s relation in RTA \cite{Cutler:1969p3655}. 
Figure \ref{fig:anaS} shows the resulting thermopower under tensile and compressive strain for a small (dashed line) 
and a five times larger charge carrier concentration (solid line).
For small charge carrier concentration the thermopower rapidly drops to constant values of 48\% for tensile and 
76\% for compressive strain, respectively. We note that the maximum of $S$ which is increased by 4\% compared to the 
unstrained case is not located at zero strain and is shifted to very 
small values of compressive strain. This behavior is more pronounced for the large charge carrier concentration, where 
the maximum of the thermopower becomes wider and is shifted to values of -0.25\% compressive strain. Again the enhancement 
reached for $S$ is about 4\% and the asymptotic values keep unchanged with doping. 
We note that for the calculation of the thermopower the influence of the effective masses 
completely cancels. For this reason neither for the \textit{ab initio} nor the analytic calculation a noticeable difference of the in-plane and 
cross-plane thermopower was found. The different saturation values of the thermopower under tensile and 
compressive strain again however can be explained in terms of a redistribution of electrons. For free electrons at T=0 
the electrical conductivity $\sigma \propto E^{3/2}$ and the resulting thermopower $S \propto E^{-1}$. 
The amount of charge carriers 
in one spin band is given by $N \propto E^{3/2}$. With \textsc{Mott}s relation $S \propto N^{-2/3}$. As stated before 
in this paper under sufficient tensile strain only the twofold degenerated $\Delta_2$ pockets are occupied compared to 
the sixfold degenerated CBM pockets in the unstrained case, so the occupation of every pocket increases by a factor of 3. One directly yields 
$\nicefrac{ S_{\text{tens.}}}{S_{0}}=(\frac{6}{2})^{-2/3}=0.48$. For compressive strain only the fourfold degenerated 
$\Delta_4$ pockets contribute, so it is $\nicefrac{ S_{\text{comp.}}}{S_{0}}=(\frac{6}{4})^{-2/3}=0.76$. The fact, that for larger 
charge carrier concentration larger strain has to be applied to reach these limits, is linked to the fact that higher strain is required to reach 
a complete redistribution of states into either the $\Delta_2$ or $\Delta_4$ pockets. 
As consequence of the discussed results we see in figure~\ref{fig:ndop}(c) an enhancement of the powerfactor in 
cross-plane direction up to 27\% at small compressive strain, while the in-plane powerfactor is only marginally 
enhanced by about 5\% under low tensile strain. 
In-plane transport under tensile strain at low doping and low temperature plays an important role in silicon based devices. 
Within figure~\ref{fig:ndop}(c) it is obvious that the strain induced influence of the powerfactor on this transport will play 
a minor role.
We want to point out that the results on the thermopower discussed above are generally valid for all systems with 
degenerate occupied bands. Lifting of the degeneracy causes redistribution of electrons and reduction of 
the thermopower. 
While the low temperature and low doping case was convenient to provide some general findings on an analytical level, 
we will now focus on the high temperature and high doping regime (see figure \ref{fig:ndop} left panels) where the power factor might be suitable 
for thermoelectric application (see also \ref{fig:unstrained}). 
At a temperature of 900K the electronic bandstructure on a width of at least $\pm \Delta E = \pm \unit[770]{meV}$ 
around the position of the chemical potential has to be included, which makes a description of the 
electronic transport properties within a spherical band picture 
impossible. Rather than providing analytical quantities a more qualitative description of our \textit{ab initio} calculations 
will be given instead. 
The electrical conductivity in figure\ref{fig:ndop}(d) states the same qualitative tendencies for \cip and \cpp 
as derived for the low temperature case. As a consequence of the high temperature and the related broadening of the Fermi-Dirac distribution 
in equations \ref{Tcoeff} and \ref{Dop} as well as the high charge carrier concentration the redistribution of states as 
described before is not completed for the strain values considered here. The analytical limits for the enhancement of \cip and \cpp for the given high doping and 
temperature should be achieved for biaxial strains of at least $\nicefrac{\Delta a}{a_{0}}=\pm 3\%$, respectively. It is 
worth mentioning, that the absolute value (cf. right scales in figures ~\ref{fig:ndop}(a),(d)) of the electrical conductivity raised enormously compared to the low doping 
case as expected. 
As a consequence, the powerfactor rises absolutely but unfortunately no enhancement via strain was obtained. 
The strain-dependent behavior of the thermopower as shown in figure \ref{fig:ndop}(e) compensates the behavior of the electrical conductivity. 
In the high doping-high temperature regime the thermopower shows a noticeable anisotropy between in-plane and cross-plane component. 
While the in-plane component \Sip confirms our analytical predictions for high doping (see black solid line in figure \ref{fig:anaS}) and even shows the 
shifted maximum to compressive strain, the cross-plane component \Spp does not follow the analytical model.
This might be explained by multiband effects and nonparabolic bands, with the latter being more relevant in the 
cross-plane direction. The overall resulting powerfactor summarized in figure \ref{fig:ndop}(f) shows however no significant anisotropy. Furthermore no 
enhancement by biaxial strain could be obtained, in contrast a decrease of about 20\% occurs.

\subsubsection{\label{strain-hole} Hole doping}
\begin{figure}[t]
\centering
\includegraphics[width=0.78\textwidth]{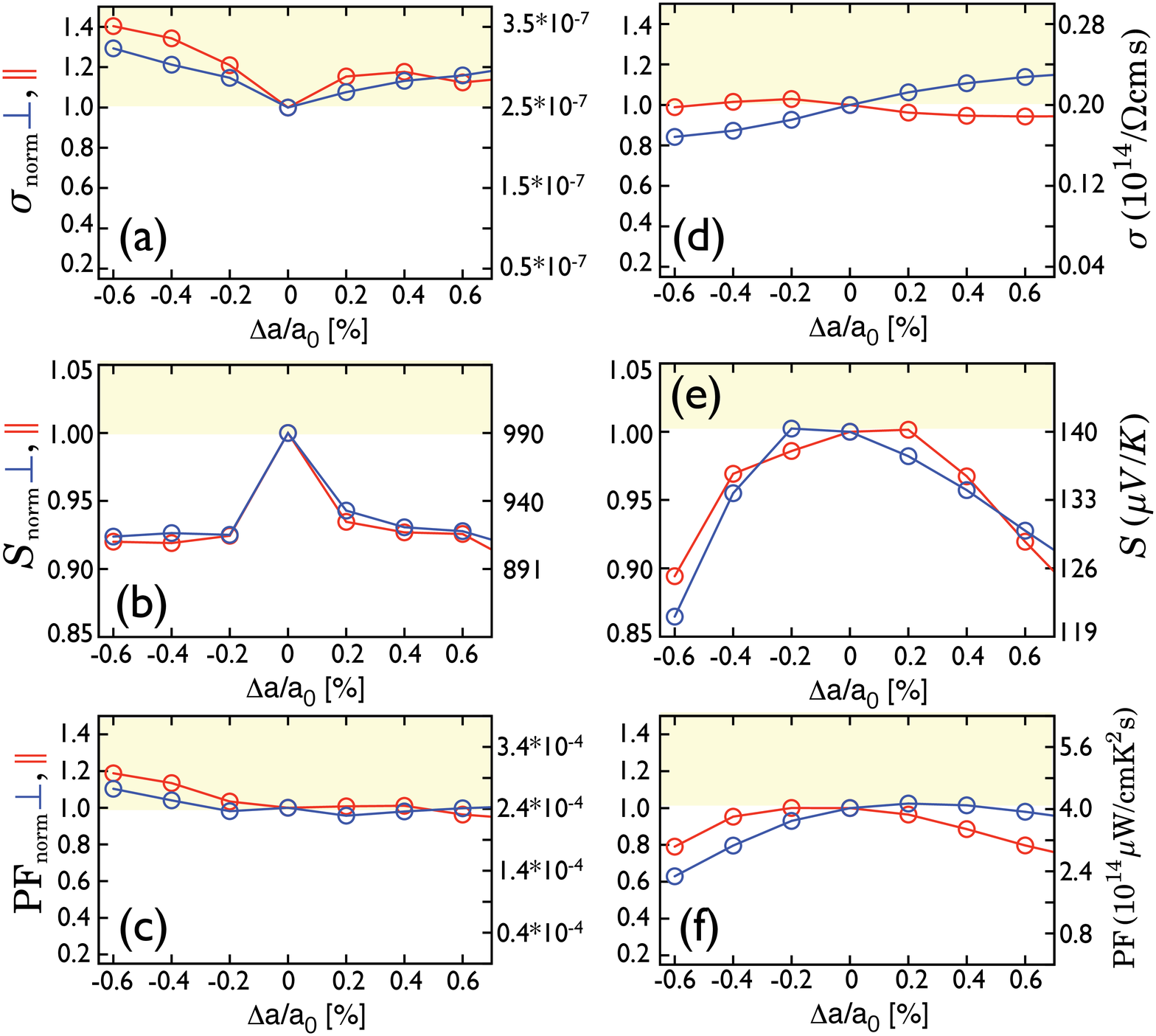}
\caption{(color online) Anisotropic thermoelectric transport properties for fixed temperature 
and hole doping concentrations in dependence on compressive and tensile strain. Left panels (a)-(c)) 
correspond to an hole doping of $\unit[1.6\cdot 10^{-7}]{h/uc}$ 
at a temperature of 100~K, while right panels refer to an hole doping of \unit[0.16]{h/uc} 
at a temperature of 900~K. On the left axis of each figure the relative value compared to the 
unstrained case is shown, while on the right axis absolute values are given. Electrical conductivity ((a),(d))
 and powerfactor ((c),(f)) are presented in units of the relaxation time $\tau$.}
\label{fig:pdop}  
\end{figure}
We will now focus in more detail on the thermolectric properties of hole doped biaxially strained 
silicon as presented in figure \ref{fig:pdop}. 
Under full relativistic 
treatment the three valence bands next to the VBM at the $\Gamma$-point are the 
heavy hole (HH), light hole (LH), and spin-orbit split-off 
(SO) hole. While the HH and LH are degenerate, the 
SO lies 44 meV apart (see figure \ref{fig:fermisur} middle panel). 
Under biaxial tensile or compressive strain, the valence bands become highly 
anisotropic and a crossover between bands occurs so that they even loose their original meaning\cite{Yu:2008p14181}. 
It was shown that mechanical deformation-induced changes in the band structure offers 
potential for significant enhancement of the hole mobility \cite{Nayak:2011p14973}.
Nevertheless, a straightforward explanation of the \textit{ab initio} calculations 
as done for the electron doped case is not any more possible.
Actually tensile and compressive biaxial strain does not only cancel the degeneracy of the heavy 
and light hole bands, which will cause reduced intervalley phonon scattering, furthermore it 
also leads to a smaller effective conductivity mass and a further depletion of the uppermost hole band 
\cite{Baykan:2011p14974,Sun:2007p14975}.
Similarly to the electron doped case in figure \ref{fig:pdop} the thermoelectric transport properties for 
hole doped silicon under the influence of biaxial strain are shown for a fixed low doping, low temperature in figure \ref{fig:pdop}(a)-(c) 
and fixed high doping, high temperature regime in figure \ref{fig:pdop}(d)-(f). 
At low temperature and slight doping an increase of the electrical conductivity was found for tensile as well as for 
compressive strain for the in-plane component \cip and the cross-plane component \cpp, while biaxial compressive 
strain tends to favor the enhancement of $\sigma$. As shown in figure \ref{fig:pdop}(b) the thermopower for hole-doped 
silicon again experiences a drop of nearly 7\% under compressive and tensile strain. 
A possible explanation for the almost symmetric drop of the thermopower under compressive and tensile strain might be 
again related to the number of bands being occupied. At small doping for compressive and for tensile biaxial strain the 
primarily occupied HH and LH split and only the upper hole band 
is depleted and dominates the character of the transport properties\cite{Baykan:2011p14974,Sun:2007p14975}. Extending 
our analytical findings for the electron doped case one would expect, that 
$\nicefrac{ S_{\text{tens./compress.}}}{S_{0}}=(\frac{2}{1})^{-2/3}=0.63$. 
Through the counteracting behavior of electrical conductivity and thermopower under strain, again no 
enhancement of the powerfactor could be found (cf. figure \ref{fig:pdop}(c)). Only under strong 
compressive biaxial strain a significant enhancement is visible in the low temperature/ low doping case.
In the high doping and temperature regime not only the upper hole band plays an important role in 
transport, furthermore the former HH, LH and SO have to be mentioned. As shown in figure \ref{fig:pdop}(d) 
an enhancement of around 
10\% compared to the unstrained case can be found for the cross-plane component \cpp under small 
tensile strain. For the in-plane electrical conductivity \cip under compressive strain only a marginal 
influence on the strain can be reported. The thermopower shows a small anisotropy of the in-plane 
and cross-plane components. While the thermopower is mainly decreased for compressive or 
tensile strain, we again see a broadening of the thermopower drop in dependence on the applied strain 
with respect to the low doping regime. 
Beside the absolute values of the powerfactor being  around 30\% smaller than in the electron doped case (comp. figure \ref{fig:ndop}(f) and 
\ref{fig:pdop}(f)), a 
slight enhancement under thermoelectrically relevant doping and temperature conditions 
could be found for the cross-plane powerfactor \PFpp under small tensile strain (see figure \ref{fig:pdop}(f)).

\subsection{\label{optpf} Optimization of the powerfactor by doping}

\begin{figure}[t]
\centering
\includegraphics[width=0.68\textwidth]{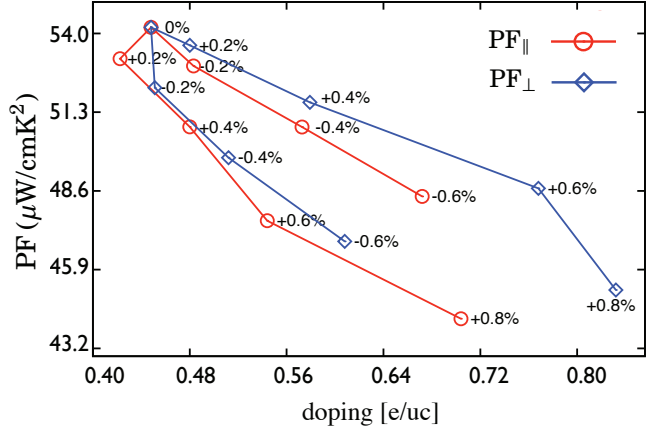}
\caption{\label{fig:6}(color online) Anisotropic power factor optimized by the carrier concentration at
given strain state for electron-doped silicon. The in-plane (cross-plane) powerfactors are drawn as red circles 
(blue diamonds). Lines are shown to guide the eyes. 
The temperature is fixed at \unit[900]{K}.}
\label{fig:nopt}  
\end{figure}

\begin{figure}[t]
\centering
\includegraphics[width=0.68\textwidth]{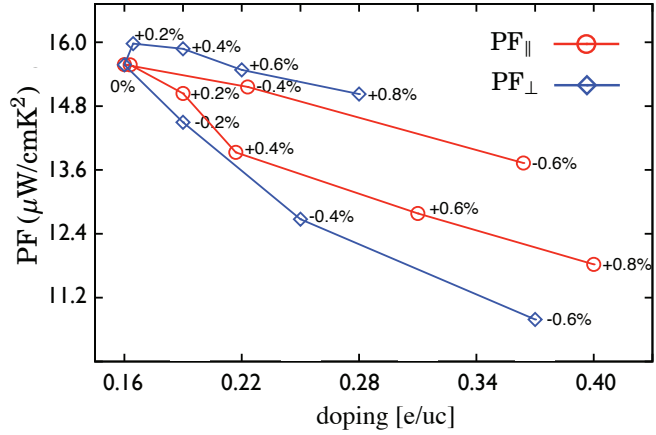}
\caption{\label{fig:7}(color online) Same as figure \ref{fig:nopt} but optimized for hole doping.}
\label{fig:popt}  
\end{figure}

While up to this point the powerfactor and the incorporated thermoelectric transport properties 
were studied for fixed temperature and charge carrier concentration in dependence on the 
applied biaxial strain, we want to gain further insight in the doping dependence. Therefore 
the amount of charge carrier concentration to optimize the powerfactor \PFip and \PFpp at given strain and fixed temperature of 900K
was determined. This temperature seems to be a common temperature for thermoelectric 
application of silicon based devices. Figures \ref{fig:nopt} and \ref{fig:popt} represent the results for the electron and hole doped 
case, respectively.

From figure \ref{fig:nopt} it becomes clear, that an enhancement of the powerfactor by sufficiently high electron doping 
can't be stated. With increasing biaxial strain the in-plane and cross-plane thermopower decreases compared 
to the unstrained case. It is worth mentioning, that the charge carrier concentration has to be increased for increasing 
tensile and compressive biaxial strain to achieve the optimal powerfactor under the certain strain  condition. Nevertheless, 
even for a raised optimal charge carrier concentration the powerfactor drops to about 80\% of the value of the unstrained 
case for the largest strain values considered here. As an interesting fact one can see that under tensile strain the 
cross-plane powerfactor \PFpp is always larger than the in-plane component \PFip, while under compressive strain it 
is the other way around and \PFpp is smaller than \PFip.
In contrast to heavily electron-doped silicon an enhancement of the powerfactor could be found for hole doped silicon, 
as shown in figure \ref{fig:popt}. 
Nevertheless, the enhancement is limited to the cross-plane contribution \PFpp. Here we find an enhanced \PFpp under 
small tensile strain of $\nicefrac{\Delta a}{a_{0}}=0.2\% \dots 0.4\% $, while for tensile strain of 0.6\% a value of the cross-plane 
powerfactor similar to the unstrained case is reached. We note, that the charge carrier concentrations, which are necessary 
to optimize the powerfactor in the hole doping case are about three times smaller than in the related electron doping case. 
Even though an enhancement of the cross-plane powerfactor under optimized hole doping and tensile strain can be found, 
unfortunately the absolute values of the powerfactor are sufficiently smaller than the absolute value of the \PFip and \PFpp of 
electron-doped silicon under all strain condition examined here.

\section{\label{ZT} Tendencies on figure of merit}

To evaluate tendencies of the figure of merit we are going to include and discuss experimental results for the 
lattice thermal conductivity $\kappa_{ph}$, which adds up to our calculated electronic thermal conductivity 
$\kappa_{el}$ for the total thermal conductivity $\kappa$. Involving the transport distribution function in eq.~\ref{Tcoeff} 
the electronic contribution to the thermal conductivity is calculated as
\begin{equation}
\kappa_{el}=\frac{1}{T}(\mathcal{L}^{(2)}-\frac{(\mathcal{L}^{(1)})^2}{\mathcal{L}^{(0)}}) \, .
\end{equation}
For bulk silicon, porous silicon and thin films various temperature- and doping-dependent 
measurements are available \cite{Brinson:1970p15026,Shanks:1963p15029,Slack:2004p15133,Bux:2009p14985,Asheghi:2002p15138}. It is well known, that the lattice thermal conductivity of silicon is strongly depends on temperature. 
While at room temperature a value of $\unit[87]{\nicefrac{W}{m \, K}}$ was reported, this value decreases to $\unit[36]{\nicefrac{W}{m \, K}}$
at $\unit[900]{K}$ and $\unit[23]{\nicefrac{W}{m \, K}}$ at $\unit[1200]{K}$ \cite{Bux:2009p14985,Shanks:1963p15029} making silicon an 
high temperature thermoelectric. 
Even at high temperatures and applicable doping the electronic contribution to the thermal 
conductivity is only a few percent of the total thermal conductivity. 
For bulk silicon it is reported, that the heat conduction is impeded by higher dopant concentrations \cite{Slack:2004p15133}. Here the lattice thermal conductivity
reduction due to the scattering by dopant ions overbalances the increase of the electronic thermal
conductivity. Due to this behaviour the total thermal conductivity $\kappa$ is weakly depends on doping for temperatures clearly above $\unit[300]{K}$ \cite{Brinson:1970p15026,Asheghi:2002p15138}.
Nevertheless, for nanostructured silicon, e.g. in silicon-based superlattices \cite{Lee:1997p1545,BorcaTasciuc:2000p15132} or 
nanoparticle bulk silicon \cite{Bux:2009p14985}, with a low lattice thermal conductivity of $\kappa_{ph}<\unit[3]{\nicefrac{W}{m \, K}}$, 
it may occur that the electronic
thermal conductivity contributes remarkably to the total thermal conductivity. At high doping levels above $\unit[1\cdot 10^{20}]{cm^{-3}}$ and temperature of $\unit[900]{K}$ the electronic contribution $\kappa_{el}$ can be around $\unit[1-3]{\nicefrac{W}{m \, K}}$ and therefore approx. $25-50\%$ of the total thermal conductivity.
While it was recently shown \cite{Li:2010p13176,Lee:1997p1545}, that tensile strain could lead to a reduction of the thermal 
conductivity by up to 15\% in bulk silicon as well as in silicon thin films we did not include this in our estimation. 
\begin{figure}[t]
\centering
\includegraphics[width=0.78\textwidth]{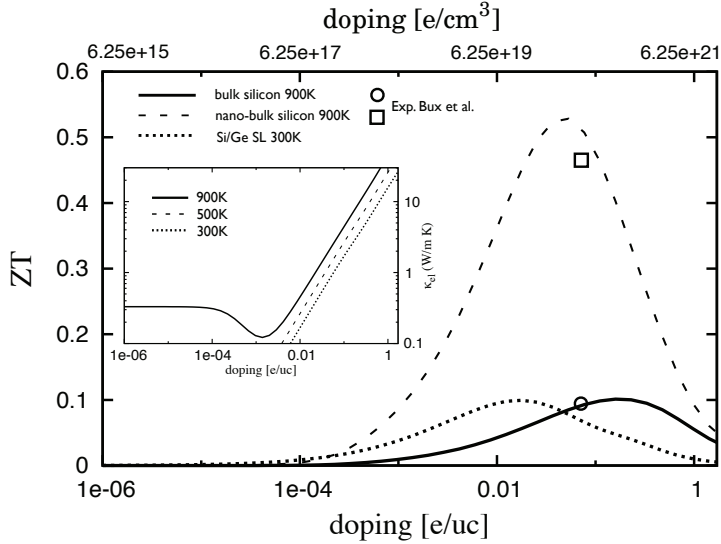}
\caption{\label{fig:9} Calculated doping-dependent figure of merit for n-doped bulk silicon at 900K (solid line), nano-bulk silicon at 900K (dashed line) and strained 
$Si/Ge$ superlattice at 300K (dotted line). The inset shows the doping-dependent electronic contribution to the total thermal conductivity for 300K, 500K and 900K. Experimental data (open square and circle) from ref.~\cite{Bux:2009p14985} are given for comparison. Note the logarithmic axes.}
\end{figure}
In figure \ref{fig:9}(a) the doping dependent figure of merit for three silicon based systems is shown. We assumed doping independent experimental 
lattice thermal conductivities stated for high doping rates, to include the reduction of $\kappa_{ph}$ by ion scattering. 
Values of $\kappa_{ph}=\unit[36]{\nicefrac{W}{m \, K}}$ \cite{Bux:2009p14985}, $\kappa_{ph}=\unit[3.8]{\nicefrac{W}{m \, K}}$\cite{Bux:2009p14985} and $\kappa_{ph}=\unit[1.9]{\nicefrac{W}{m \, K}}$\cite{Lee:1997p1545} were used for 
single crystalline bulk silicon, nanostructured bulk silicon and a $Si/Ge$ superlattice, respectively. Here we consider only electron doping, which appeared to be most promising. 
The calculated doping dependent electronic  part $\kappa_{el}$ is shown in the inset of fig.~\ref{fig:9} for three different temperatures. For bulk silicon (solid line in fig.~\ref{fig:9}) we find a broad maximum for the figure of merit $ZT \approx 0.1$ at 900K and high doping levels of $N \approx \unit[6.5 \dots 18 \cdot 10^{20}]{cm^{-3}}$, which is in good agreement to experiments by Bux \textit{et al.}~\cite{Bux:2009p14985}. Compared to the related powerfactor (see fig.~\ref{fig:1}) the maximum shifts to lower doping concentrations due to the linear increase of $\kappa_{el}$ at higher doping rates, and therefore a stronger decrease of ZT at high doping rates. If a reduction of the lattice thermal conductivity can be achieved by nano inclusions as reported in ref.~\cite{Bux:2009p14985} a remarkable $ZT$ of about $0.5$ at $N=\unit[3 \cdot 10^{20}]{cm^{-3}}$ at 900K could be obtained making silicon an interesting high temperature thermoelectric (dashed line in fig.~\ref{fig:9}). Nevertheless, at room temperature the lowest thermal conductivities have been stated for silicon and germanium based superlattices \cite{Lee:1997p1545,BorcaTasciuc:2000p15132}. Therefore the dotted line fig.~\ref{fig:9} estimates the figure of merit for a Si/Ge superlattice with a period 
of $\unit[15]{nm}$ assuming no degradation of the electronic transport by the heterostructure. Here, at 300K, the maxium figure of merit is found around doping levels of $\unit[7 \cdot 10^{19}]{cm^{-3}}$ with values of about $0.1$, which are comparable to single crystalline silicon at 900K at ten times larger doping levels.

\section{Conclusion} 
In conclusion, the thermoelectric transport properties of biaxially strained silicon 
were studied in detail with respect to a possible enhancement of the powerfactor. Two different doping and temperature 
regimes were analyzed in detail. A low doping ($N \approx \unit[1\cdot 10^{14}]{cm^{-3}}$) and low temperature regime (T=100K) 
suitable for metal-oxide-semiconductor device applications and a heavy doping ($N \approx \unit[1\cdot 10^{20}]{cm^{-3}}$) and 
high temperature regime (T=900K) suitable for silicon based thermoelectric modules.

It was shown that the 
electronic transport properties, namely the electrical conductivity $\sigma$ and the thermopower $S$ are highly sensitive against strain. 
Nevertheless it was found that strain-induced effects in $\sigma$ and $S$ compensate each other 
and no remarkable enhancement of the powerfactor by either compressive or tensile biaxial strain can be reached. On the other hand a reduction 
of the powerfactor of upto 20\% (30\%) under electron (hole) doping due to the influence of strain was found. The latter was 
assigned to a large extent to band-structure related redistribution of electrons. Estimations for the figure of merit under electron doping are given.

As a general result we showed that when degenerate VBM or CBM exist the thermopower decreases with deformation due to 
a redistribution of electrons in energetically lifted valleys. This could be the general explanation of the reduction of 
the electronic thermoelectric properties in strain-influenced heterostructures.

\begin{ack}
  This work was supported by the Deutsche
  Forschungsgemeinschaft, SPP 1386 `Nanostrukturierte Thermoelektrika: 
  Theorie, Modellsysteme und kontrollierte Synthese'. N. F. Hinsche is
  member of the International Max Planck Research School for Science
  and Technology of Nanostructures. We want to thank 
  Florian Rittweger for computational assistance.

\end{ack}

\section*{References}
\bibliography{sub_1_revpz.bbl}

\end{document}